# An Integrated Framework of Spatial Autocorrelation Analysis Based on Gravity Model

Yanguang Chen

(Department of Geography, College of Urban and Environmental Sciences, Peking University, 100871, Beijing, China. Email: chenyg@pku.edu.cn)

**Abstract**: Spatial interaction and spatial autocorrelation are two different fields of geo-spatial analysis, revealing the internal relationship between the two fields will help to develop the theory and method of geographical analysis. This paper is devoted to deducing a system of spatial correlation analysis models from the gravity model by mathematical derivation. The main results are as follows. First, a set of potential energy measurements are derived from the gravity model. Second, a pair of correlation equations, including an inner product equation and an outer product equation, are constructed based on the quadratic form of potential energy formula. Third, a series of spatial autocorrelation statistics, including Moran's index and Getis-Ord's index are derived from the potential energy formula. Fourth, the concept of fractal dimension is introduced into spatial weight matrix. The observational data of urban systems are employed to make an empirical analysis, demonstrating the application procedure of newly derived models. A conclusion can be drawn that spatial autocorrelation is actually rooted in spatial interaction process, and an improved methodology of spatial analysis can be developed by integrating spatial autocorrelation models and gravity model into the same framework.

## 1 Introduction

The gravity model of urban and regional systems by analogy with Newton's gravitation is one of the traditional tools in geographical spatial analysis. The model describes the human force of attraction between any two places, which is directly proportional to the product of their sizes and inversely proportional to the square of the distance between them. Based on the gravity model indicating attraction force, the potential energy concept was originally derived for social physics by Stewart (1948; 1950a; 1950b), a Princeton University astrophysicist. In the period of quantitative revolution of geography (1953-1976), the measurements of demographic force, energy, and potential were introduced into human geography (James and Martin, 1981; Martin, 2005). The distance decay effect of geographical actions and distributions was verified by observational data, and the gravity model is empirically effective (Stewart, 1942; Haggett *et al*, 1977; Chen, 2015). Because of ubiquitous distance decay phenomena, Tobler (1970; 2004) presented the first law of geography which reads "everything is related to everything else but near things are more related than distant things". The distance decay function is originally an inverse power function, which suggests proportional relations and dimensional consistency. However, the observed values of the distance exponent cannot be interpreted with the dimension concept from Euclidean geometry (Haynes, 1975). This caused a dimension puzzle of the gravity model (Chen, 2015). As an alternative of the inverse power law, the negative exponential was employed to serve for the distance decay function (Haggett *et al*, 1977; Wilson, 2010). However, the exponential distance decay suggests spatial localization rather than action at a distance and thus defies the first law of geography (Chen, 2015). As a result, the gravity model and potential formula fell in a theoretical dilemma.

A new finding is that the potential analysis can be associated with spatial autocorrelation analysis. Local Getis-Ord's index was proved to be a normalized potential index (Chen, 2020). Both the potential model and the spatial autocorrelation model belong to the system of spatial correlation models. In urban studies, there is an analogy between the Moran's index and the sum of the potential energy index of a system of cities. The theory and methods of spatial autocorrelation analysis have been welled developed (see e.g., Bivand *et al*, 2009; Cliff and Ord, 1973; Cliff and Ord, 1981; Fischer and Wang, 2011; Getis, 2009; Griffith, 2003; Haining, 2009; Odland, 1988; Sokal and Oden, 1978; Sokal and Thomson, 1987; Wang, 2012). In contrast, the potential energy model and potential-based spatial analysis have not evolved into a logic system. The reason for this is that the relation between spatial autocorrelation model and the dimension conundrum is not clear, while there is a clear relation between gravity model and the dimension puzzle. Fortunately, the power-law distance decay function has gone out of its dimension dilemma because of fractal geometry. The distance exponent can be efficiently interpreted with the idea from fractal dimension (Chen, 2015).

Today, it is time to derive generalized spatial autocorrelation models from the gravity models. The potential energy model and potential analysis can be reconstructed and further advanced by analogy with spatial autocorrelation theory. This paper is devoted to developing a new framework of spatial correlation analysis based on the classical models such as the gravity model and potential energy formula. Thus the spatial interaction and spatial autocorrelation analysis can be integrated into a general framework. The rest of this paper is organized as follows. In Section 2, the traditional potential formula are extended to form a systematic analytical process, which comprises global measurements, local measurements, and scatterplots. The underlying rationale is demonstrated by mathematical reasoning. The main measurements are put in order, a comparison between the new spatial correlation models and the spatial autocorrelation models is drawn. In Section 3, as a case study, the newly advanced methodology is applied to capital cities of China to show how to make a spatial correlation analysis based on newly developed potential energy models. In Section 4, several related questions are discussed, and finally, in Section 5, the discussion is concluded by summarizing the main points of this work.

## 2 Theoretical Results

### 2.1 Global potential and local mutual energy

Potential index is frequently applied to the spatial analysis of urban systems, thus this theoretical study is empirically based on the concept of cities. The final results can be generalized to the related fields such as demography, regional geography, and spatial economics. Suppose that there is a system of $n$ cities in a geographical region. The size of each city can be measured with urban population, and population size reflects the first dynamics of cities (Arbesman, 2012). By analogy with Newton's law of universal gravitation, the gravity between any two cities can be expressed as (Haggett, 2001; Rybski *et al*, 2013; Taylor, 1983; Wilson, 2000)

$$F_{ij} = K\frac{Q_iO_j}{r_{ij}^b}, \tag{1}$$

where $F_{ij}$ denotes the "attraction" between the $i$th city and the $j$th city ($i$, $j$=1, 2,…, $n$), $r_{ij}$ is the distance between cities $i$ and $j$, $Q_i$ and $Q_j$ are the "masses" (sizes) of the two cities, $K$ refers to the gravity coefficient, and $b$, which comes between 0 and 3, to the decay exponent indicating friction of distance (Haggett *et al*, 1977). The model states that the gravity between two cities is proportional to the product of the two cities' sizes and inversely proportional to the $b$th power of the distance between the two cities. In theory, based on allometric scaling relations, equation (1) proved to be fractal model and can be derived from the revised spatial interaction model (Chen, 2015), which is based on the principle of entropy-maximization (Wilson, 1968; Wilson, 2010). The distance decay exponent is a fractal parameter associated with Zipf's law and central place networks (Chen, 2011).

The gravity model describes the human force of attraction between any two urban places, but it cannot be used to reflect the attractive effect between the $n$ cities. By analogy with the concept of potential energy in Newtonian physics, Stewart (1950) proposed a potential formula to measure the attractive strength between a city and the other $n$-1 cities. The potential energy of city $i$ and city $j$ can be defined by

$$E_{ij} = F_{ij} r_{ij} = K \frac{Q_i Q_j}{r_{ij}^{b-1}} . \tag{2}$$

If the parameter values are $K$=1 and $b$=2, then the result is just the model proposed by Zipf (1946). The sum of potential energy of city $i$ relative to all the other $n$-1 cities is

$$E_i = \sum_{j=1}^{n-1} E_{ij} = K Q_i \sum_{j=1}^{n-1} \frac{Q_j}{r_{ij}^{b-1}} , \tag{3}$$

which indicates a kind of local potential energy (LPE), as will be illuminated in next subsection ($i \neq j$). Thus the relative potential index of a city in an urban system can be expressed with

$$U_i = \frac{E_i}{Q_i} = K \sum_{j=1}^{n-1} \frac{Q_j}{r_{ij}^{b-1}} , \tag{4}$$

where $U_i$ refers to the relative potential index of city $i$ relative to all the other cities in the systems of cities. A high value of $U_i$ suggests a good accessibility of a city in its geographical region (Zhou, 1995).

The concept of potential energy is useful in spatial analysis of geographical systems. However, its function has limitation. A new measure of urban potential energy can be defined to develop the potential model for urban studies. Based on the LPE formula, we can define a global potential energy (GME) for a system of $n$ cities, that is

$$E = \sum_{i=1}^{n-1} E_i = K \sum_{i=1}^{n=1} \sum_{j=1}^{n-1} \frac{Q_i Q_j}{r_{ij}^{b-1}} , \tag{6}$$

where $E$ suggests a GPE of the $n$ cities ($i \neq j$). Equation (6) leaves out the relationship between a city and itself because $r_{ii}=r_{jj}=0$. In fact, we can define that $E_{ii}=E_{jj}=0$, and thus, equation (6) can be expressed as a generalized quadratic form as follows

$$E = K\sum_{i=1}^{n}\sum_{j=1}^{n}\frac{Q_i Q_j}{r_{ij}^{b-1}} = K\mathbf{Q}^{\mathrm{T}}\mathbf{V}\mathbf{Q}, \tag{7}$$

where $\mathbf{Q}^{\mathrm{T}}=[Q_1, Q_2, \ldots, Q_n]$ refers to the transpose of a city size vector $\mathbf{Q}$, and $\mathbf{V}$ to the spatial contiguity matrix (SCM), that is

$$\mathbf{V} = \left[v_{ij}\right]_{n\times n} = \begin{bmatrix} v_{11} & v_{12} & \cdots & v_{1n} \\ v_{21} & v_{22} & \cdots & v_{2n} \\ \vdots & \vdots & \ddots & \vdots \\ v_{n1} & v_{n2} & \cdots & v_{nn} \end{bmatrix}. \tag{8}$$

According to equation (7), the entries of SCM can be generated by the following contiguity function

$$v_{ij} = \begin{cases} 1/r_{ij}^{b-1}, & i \neq j \\ 0, & i = j \end{cases}. \tag{9}$$

Apparently, the GPE is similar to the generalized quadratic form of Moran's index on spatial autocorrelation (Chen, 2013). Of course, $E$ is not an actually a quadratic form because $\mathbf{V}$ is not a positive definite matrix. The LPE can be re-expressed as

$$E_i = KQ_i\sum_{j=1}^{n} v_{ij}Q_j, \tag{10}$$

which is mathematically similar to the local indicators of spatial association (LISA). LISA was proposed by Anselin (1995) and can be used for local spatial autocorrelation analysis.

The potential index can be divided into global potential index and local potential indexes. The local relative potential index (LRPI) has been defined by equation (4). Corresponding to equation (7), a local potential vector based on equation (4) can be expressed as below

$$\mathbf{U} = \begin{bmatrix} U_1 & U_2 & \cdots & U_n \end{bmatrix}^{\mathrm{T}} = K\mathbf{V}\mathbf{Q}, \tag{11}$$

which yields a set of LPI values of the $n$ cities. The global relative potential index (GRPI) is

$$U = \sum_{i=1}^{n} U_i = K\sum_{i=1}^{n}\sum_{j=1}^{n}\frac{Q_j}{r_{ij}^{b-1}} = K\sum_{i}^{n}(\mathbf{VE})_i, \tag{12}$$

in which $U$ denotes global relative potential index, and local relative potential index $U_i=0$ for $i=j$. If the city population size variable is normalized, the LRPI proved to be equivalent to Getis-Ord's indexes (Chen, 2020).

## 2.2 Matrix scaling and two correlation matrixes

The vector of city sizes has an inner product and an outer product, based on which two spatial correlation matrixes can be constructed. The inner product is a scalar:

$$\Lambda = \mathbf{Q}^{\mathrm{T}}\mathbf{Q} = [Q_1 \quad Q_2 \quad \cdots \quad Q_n]\begin{bmatrix} Q_1 \\ Q_2 \\ \vdots \\ Q_2 \end{bmatrix} = \sum_{i=1}^{n} Q_i^2 , \tag{13}$$

which is also termed dot product or scalar product. The outer product is a matrix:

$$\Omega = \mathbf{Q}\mathbf{Q}^{\mathrm{T}} = \begin{bmatrix} Q_1 \\ Q_2 \\ \vdots \\ Q_n \end{bmatrix}[Q_1 \quad Q_2 \quad \cdots \quad Q_n] = \begin{bmatrix} Q_1Q_1 & Q_1Q_2 & \cdots & Q_1Q_n \\ Q_2Q_1 & Q_2Q_2 & \cdots & Q_2Q_n \\ \vdots & \vdots & \ddots & \vdots \\ Q_nQ_1 & Q_nQ_1 & \cdots & Q_nQ_n \end{bmatrix} . \tag{14}$$

It is easy to prove that the inner product $\Lambda$ is just the eigenvalue of the outer product $\mathbf{\Omega}$ corresponding to the eigenvector $\mathbf{Q}$. The proof is as below:

$$\mathbf{\Omega Q} = \mathbf{Q}\mathbf{Q}^{\mathrm{T}}\mathbf{Q} = \Lambda\mathbf{Q} , \tag{15}$$

from which it follows a symmetric relation as follows

$$\mathbf{Q}\mathbf{Q}^{\mathrm{T}}\mathbf{Q} = \mathbf{Q}\Lambda = \Lambda\mathbf{Q} = (\mathbf{Q}^{\mathrm{T}}\mathbf{Q})\mathbf{Q} . \tag{16}$$

This relation is important for the spatial analysis based on the concepts of potential energy. Developing equation (15) yields

$$\begin{bmatrix} Q_1 \\ Q_2 \\ \vdots \\ Q_n \end{bmatrix}[Q_1 \quad Q_2 \quad \cdots \quad Q_n]\begin{bmatrix} Q_1 \\ Q_2 \\ \vdots \\ Q_n \end{bmatrix} = \begin{bmatrix} Q_1\sum_{i=1}^{n} Q_i^2 \\ Q_2\sum_{i=1}^{n} Q_i^2 \\ \vdots \\ Q_n\sum_{i=1}^{n} Q_i^2 \end{bmatrix} = \Lambda\begin{bmatrix} Q_1 \\ Q_2 \\ \vdots \\ Q_n \end{bmatrix} . \tag{17}$$

For example, for $n$=2, the extended form of equation (15) is

$$\begin{bmatrix} Q_1 \\ Q_2 \end{bmatrix}[Q_1 \quad Q_2]\begin{bmatrix} Q_1 \\ Q_2 \end{bmatrix} = \begin{bmatrix} Q_1Q_1 & Q_1Q_2 \\ Q_2Q_1 & Q_2Q_2 \end{bmatrix}\begin{bmatrix} Q_1 \\ Q_2 \end{bmatrix} = \begin{bmatrix} Q_1(Q_1^2+Q_2^2) \\ Q_2(Q_1^2+Q_2^2) \end{bmatrix} = \Lambda\begin{bmatrix} Q_1 \\ Q_2 \end{bmatrix} , \tag{18}$$

where $\Lambda=Q_1^2+Q_2^2$. Further, it can be shown that $\Lambda$ is the maximum eigenvalue of $\mathbf{\Omega}$. For a square matrix, the trace of $\mathbf{\Omega}$ is

$$\mathrm{T_r}(\Omega) = Q_1^2 + Q_2^2 + \cdots Q_n^2 = \Lambda = \lambda_1 + \lambda_2 + \cdots + \lambda_n , \tag{19}$$

where $T_r$ refers to "finding the trace (of $\Omega$)". Given $\lambda_1=\lambda_{max}=\Lambda$, it follows that

$$\lambda = \begin{cases} \Lambda, & \lambda = \lambda_{\max} \\ 0, & \lambda \neq \lambda_{\max} \end{cases}. \tag{20}$$

According to the Cayley-Hamilton theorem, the eigenvalues of any $n$-by-$n$ square matrix are identical to the roots of its corresponding polynomial equation. For example, for $n$=2, the characteristic polynomial of the matrix $\mathbf{\Omega}$ is

$$\lambda\mathbf{I} - \mathbf{\Omega} = \begin{vmatrix} \lambda - Q_1Q_1 & -Q_1Q_2 \\ -Q_1Q_2 & \lambda - Q_2Q_2 \end{vmatrix} = \lambda^2 - \lambda(Q_1^2 + Q_2^2) = \lambda^2 - \Lambda\lambda = 0, \tag{21}$$

where $\mathbf{I}$ denotes the identity/unit matrix. Thus we have

$$\lambda_1 = \sum_{i=1}^{2} Q_i^2 = \Lambda, \quad \lambda_2 = 0. \tag{22}$$

This indicates that the maximum eigenvalue of the outer product matrix $\mathbf{\Omega}$ is the corresponding inner product $\Lambda$.

In general scientific research, mathematical modeling and quantitative analysis are in fact based on characteristic scales. If and only if we can find some types of characteristic lengths or characteristic parameters, we can make effective spatial analysis. It can be proved that the GPE index is actually a characteristic value of spatial correlation matrix. In terms of equation (7), using the outer product $\mathbf{\Omega}$ to multiply equation (11) left yields

$$\mathbf{\Omega V Q} = \mathbf{Q Q}^{\mathrm{T}} \mathbf{V Q} = E\mathbf{Q}. \tag{23}$$

This suggests that the GPE index is the eigenvalue of $\mathbf{\Omega V}$ corresponding to the eigenvector $\mathbf{Q}$. In terms of equation (13), the left multiplication of equation (11) by the inner product $\Lambda$ yields

$$\Lambda\mathbf{VQ} = \mathbf{Q}^{\mathrm{T}}\mathbf{QU} = \mathbf{Q}^{\mathrm{T}}\mathbf{QVQ}, \tag{24}$$

which indicates the precondition of equation (7), that is

$$\Lambda\mathbf{VQ} = \mathbf{Q}^{\mathrm{T}}\mathbf{QVQ} = E\mathbf{Q}. \tag{25}$$

Apparently, multiplying equation (25) left with the transpose of $\mathbf{Q}$ yields

$$\mathbf{Q}^{\mathrm{T}}\Lambda\mathbf{VQ} = \mathbf{Q}^{\mathrm{T}}\mathbf{QQ}^{\mathrm{T}}\mathbf{VQ} = E\mathbf{Q}^{\mathrm{T}}\mathbf{Q} = E\mathbf{\Lambda}, \tag{26}$$

from which it follows equation (7) and gives the GPE value. According to equation (16), if the SCM is replaced with an identity matrix, equation (23) and equation (25) will be the same with each other.

A pair of important matrixes can be obtained from the above mathematical processes and results. Then two useful variables can be defined for systematic potential analysis. The first matrix is based on the outer product:

$$\mathbf{A}^{*} = K\mathbf{Q}\mathbf{Q}^{\mathrm{T}}\mathbf{V}, \tag{27}$$

which can be termed *ideal spatial correlation matrix* (ISCM). The second matrix is based on the inner product:

$$\mathbf{A} = K\mathbf{Q}^{\mathrm{T}}\mathbf{Q}\mathbf{V}, \tag{28}$$

which can be termed *real spatial correlation matrix* (RSCM). Based on equation (27), a strict matrix scaling relation can be given as

$$\mathbf{A}^{*}\mathbf{Q} = E\mathbf{Q}, \tag{29}$$

which is equivalent to equation (23). Based on equation (28), a possible matrix scaling relation can be constructed as

$$\mathbf{A}\mathbf{Q} = E\mathbf{Q}, \tag{30}$$

which is equivalent to equation (25). Using equation (29) and equation (30), we can estimate the GPE index.

The ISCM is significant for mutual energy calculation. Both the GPE and LPEs can be computed with equation (29), which can be developed as

$$\mathbf{A}^{*} = K\mathbf{Q}\mathbf{Q}^{\mathrm{T}}\mathbf{V} = K\begin{bmatrix} Q_1\sum_{i,j=1}^{n} v_{i1}Q_j & Q_1\sum_{i,j=1}^{n} v_{i2}Q_j & \cdots & Q_1\sum_{i,j=1}^{n} v_{in}Q_j \\ Q_2\sum_{i,j=1}^{n} v_{i1}Q_j & Q_2\sum_{i,j=1}^{n} v_{i2}Q_j & \cdots & Q_2\sum_{i,j=1}^{n} v_{in}Q_j \\ \vdots & \vdots & \ddots & \vdots \\ Q_n\sum_{i,j=1}^{n} v_{i1}Q_j & Q_n\sum_{i,j=1}^{n} v_{i2}Q_j & \cdots & Q_n\sum_{i,j=1}^{n} v_{in}Q_j \end{bmatrix}. \tag{31}$$

The diagonal elements of the final matrix in equation (31) proved to be the LPE of the $n$ cities, which can be expressed as equation (10). The sum of the diagonal entries gives GPE. The members of the matrix are as below

$$E_{ij} = KQ_i\sum_{i,j=1}^{n} v_{ij}Q_j\ . \tag{32}$$

The row's sums by columns is

$$J_i = \sum_{j=1}^{n} E_{ij} = K\sum_{j=1}^{n}(Q_i \sum_{i=1}^{n} v_{ij} Q_j) = KQ_i \sum_{j=1}^{n}(\mathbf{VQ})_j = UQ_i , \tag{33}$$

where $U$=∑$U_i$ denotes the GRPI. The column's sums by rows is

$$H_j = \sum_{i=1}^{n} E_{ij} = K\sum_{i=1}^{n}(Q_i \sum_{j=1}^{n} v_{ij} Q_j) = K(\mathbf{VQ})_j \sum_{i=1}^{n} Q_i = SU_j , \tag{34}$$

where $S$=∑$Q_i$ represents the sum of $Q_i$. Apparently, there is a dual relationship between equation (33) and equation (34). The $J_i$ values can be associated with $\mathbf{QQ}^{\mathrm{T}}\mathbf{VQ}$, while the $H_i$ values can be associated with $\mathbf{Q}^{\mathrm{T}}\mathbf{QVQ}$. The ratio of $J_i$ to $Q_i$ gives the GPE, $U$, which is formulated as equation (12).

## 2.3 Potential energy scatterplots

Two graphs can be employed to make absolute and relative potential energy analyses for spatial correlation and spatial interaction. One of them is similar to the Moran's scatterplot on spatial autocorrelation (Anselin, 1995; Anselin, 1996). The set of scatterplots comprises size-potential plot and absolute -relative potential plot. The potential energy scatterplot is useful for spatial potential energy analysis. In order to create the scatterplots for potential energy, two vectors can be defined as follows

$$\mathbf{f}^* = \mathbf{A}^*\mathbf{Q} = \mathbf{QQ}^{\mathrm{T}}\mathbf{WQ} = E\mathbf{Q} , \tag{35}$$

$$\mathbf{f} = \mathbf{AQ} = \mathbf{Q}^{\mathrm{T}}\mathbf{QWQ} \rightarrow E\mathbf{Q} . \tag{36}$$

Using equations (35) and (36), we can make a normalized size-potential energy plot consisting of $n$ scattered points and a trend line. The relationship between $Q$ and $f^*$ suggests the theoretical potential energy distribution, i.e., the regression line through ordered points, and the dataset of $Q$ and $f$, denotes the actual potential energy pattern, i.e. the randomly distributed points on the scatterplot. The slope of the regression line gives the GPE value. We have two ways of generating the regression line. One is to connect the ordered points of $f^*$ based on $Q$, and the other is to add a trend line by using the least squares regression of $f^*$ based on $Q$ with the constant term being zero. This indicates that the trend line based on the ordered points and the trend line based on the scattered points overlap one another. In urban studies, the potential energy plot can be used to reflect the relative significance of a city within the urban systems visually.

The relative potential scattered plot can also be used to make a potential analysis. This plot can be generated by equations (33) and (34). Using $Q$ to represent the $x$-coordinate, and $J$ and $H$ to represent the $y$- coordinate, we can make another potential energy scatterplot. The relationships between $J$ and $Q$ shows a set of ordered data points, which yields a straight line, while the relationships between $H$ and $Q$ displays a set of randomly scattered data points. The slope of the regression line based on the ordered points gives the GRPI, $U$. However, if we add a trend line to the scattered points by means of the least squares regression with the constant term being zero, the slope is not equal to the GRPI value. In other words, the trend line based on the ordered points and the trend line based on the scattered points do not overlap each other. They form an angle greater than 0. By the way, the relation between $J_i$ and $H_i$ yields a scatterplot, which can be utilized to replace the size-potential scatterplot defined above. However, the slope of the trend line is not equal to the GPE.

The size-potential plots and the relative potential plot can be employed to reveal visually the relative importance of a city in a system of cities. In fact, we can define the third scatterplot, absolute-relative potential plot. Using the relative potential index as the horizontal axis and the absolute potential index as the vertical axis, it is easy to create this scatter plot. This scatter plot can be used to visually reflect the comprehensive importance of a city in the urban system.

### 2.4 Link to spatial autocorrelation

A comparison can be drawn between the potential-based spatial correlation model and the common spatial autocorrelation models. The potential energy index is to Moran's index what covariance is to Pearson's correlation coefficient in statistics. The simple correlation coefficient presented by Pearson can be treated a "standardized" covariance of two variables (Moore, 2009). Similarly, the spatial autocorrelation coefficient proposed by Moran (1948) can be regarded as a "standardized" potential index. In order to complement the function of Moran's index, Getis and Ord (1992) proposed new indexes. From the potential energy indexes, we can derive Getis-Ord's indexes and Moran's indexes. Suppose that the original size variable is $x$, and the corresponding vector is x=$[x_1, x_2,\ldots, x_n]^{\mathrm{T}}$. The size vector $Q$ may be a normalized variable based on the sum of $x$, and the formula is as below:

$$Q=\frac{x}{X}=\frac{x}{\sum_{i=1}^{n}x_i}=p\,, \tag{37}$$

in which $X=\sum x_i$ denotes the sum of $x_i$, and $p$ refers to a normalized variable based on the sum of $x$. In equation (37), the diagonal elements of the outer product matrix of $\mathbf{X}$ is taken into account. If the diagonal elements of the outer product matrix is removed, equation (37) should be revised as

$$Q=\frac{x}{X^{*}}=\frac{x}{\sqrt{(\sum_{i=1}^{n} x_i)^2-\sum_{i=1}^{n} x_i^2}}=p^{*}, \tag{38}$$

where $X^{*}=((\sum x_i)^2 - \sum x_i^2)^{1/2}$. On the other hand, the size vector $\mathbf{x}$ can be standardized by

$$Q=\frac{x-\mu}{\sigma}=z, \tag{39}$$

where $\mu$ refers to the average value of $x$, $\sigma$ to the corresponding population standard deviation (PSD), and $z$ denotes $Z$-score. The SCM can be normalized by sum and converted into a spatial weight matrix (SWM), that is

$$\mathbf{W}=K\mathbf{V}=\frac{\mathbf{V}}{V_0}=\begin{bmatrix} w_{11} & w_{12} & \cdots & w_{1n} \\ w_{21} & w_{22} & \cdots & w_{2n} \\ \vdots & \vdots & \ddots & \vdots \\ w_{n1} & w_{n2} & \ddots & w_{nn} \end{bmatrix}, \tag{40}$$

where

$$V_0=\sum_{i=1}^{n}\sum_{j=1}^{n} v_{ij}=\frac{1}{K},\quad w_{ij}=\frac{v_{ij}}{V_0},\quad w_{ii}=w_{jj}=0,\quad \sum_{i=1}^{n}\sum_{j=1}^{n} w_{ij}=1. \tag{41}$$

Without loss of generality, we can define $V_0=k/K$, where $k$ denotes a coefficient reflecting spatial diffusion. But for comparability and simplicity, let $k=1$. Substituting the original size vector and SCM in equation (7) with sum-based normalized vector and SWM yields the global Moran's index, which can be calculated by (Chen, 2013)

$$I=\mathbf{Q}^{\mathrm{T}}\mathbf{W}\mathbf{Q}=(\frac{x-\mu}{\sigma})^{\mathbf{T}}\frac{\mathbf{V}}{V_0}(\frac{x-\mu}{\sigma})=\mathbf{z}^{\mathbf{T}}\mathbf{W}\mathbf{z}, \tag{42}$$

where $\mathbf{z}=[z_1, z_2,\ldots, z_n]^{\mathrm{T}}$. Replacing the size vector and SCM in equation (7) by the sum-based normalized vector and SWM yields the global Getis-Ord's index, that is

$$G=\mathbf{Q}^{\mathrm{T}}\mathbf{W}\mathbf{Q}=(\frac{\mathbf{x}}{X})^{\mathbf{T}}\frac{\mathbf{V}}{V_0}(\frac{\mathbf{x}}{X})=\mathbf{p}^{\mathbf{T}}\mathbf{W}\mathbf{p}, \tag{43}$$

where $\mathbf{p}=[p_1, p_2,\ldots, p_n]^{\mathrm{T}}$. If the diagonal elements in the outer product of $\mathbf{Q}$ are not considered, equation (43) should be substituted by

$$G^{*}=\mathbf{Q}^{\mathrm{T}}\mathbf{W}\mathbf{Q}=(\frac{\mathbf{x}}{X^{*}})^{\mathrm{T}}\frac{\mathbf{V}}{V_{0}}(\frac{\mathbf{x}}{X^{*}})=\frac{1}{1-\mathbf{p}^{\mathrm{T}}\mathbf{p}}\mathbf{p}^{\mathrm{T}}\mathbf{W}\mathbf{p}=\mathbf{p}^{*\mathrm{T}}\mathbf{W}\mathbf{p}^{*}. \tag{44}$$

The normalized local Moran's indexes can be given by the diagonal elements of the follows matrix

$$\mathbf{M}^{*}=\mathbf{z}\mathbf{z}^{\mathrm{T}}\mathbf{W}. \tag{45}$$

The local Getis-Ord's indexes can be computed by the formula as below

$$\mathbf{G}=\mathbf{W}\mathbf{p},\quad \mathbf{G}^{*}=\mathbf{W}\mathbf{p}^{*}. \tag{46}$$

Thus the internal logical relations between potential-based interaction analysis and spatial autocorrelation analysis are brought to light.

In light of the mathematical process of derivation of Getis-Ord's indexes and Moran's indexes from the potential-based spatial correlation index, we can find the similarities and differences between the new spatial correlation models advanced in this work and the well-known spatial autocorrelation measures. The global potential energy index is equivalent to Getis-Ord's index and identical in form to Moran's index (Table 1). The global Getis-Ord's index is actually rescaled GPE index. Differing from Moran's index which is based on dimensionless variables, the potential-based indexes contain the information of size measurements and spatial distances. Thus Moran's indexes can be used to reveal relative strength of spatial correlation, while the potential-based measurements and the potential energy index can be employed to reflect the both absolute and relative strengths of spatial correlation and interaction. Where functions are concerned, Getis-Ord's indexes come between Moran's indexes and energy indexes. Based on the potential-based correlation model, the gravity model and the spatial autocorrelation can be integrated into a general framework of spatial correlation analysis (Figure 1).

**Table 1 The similarities and differences between the spatial correlation model based on potential energy and common spatial autocorrelation models**

| **Item** | | **Spatial correlation model** | **Spatial autocorrelation model** | |
|---|---|---|---|---|
| | | | Getis-Ord's $G$ | Moran's $I$ |
| **Global index** | | $E=\mathbf{Q}^{\mathrm{T}}\mathbf{V}\mathbf{Q}$ | $G=\mathbf{p}^{\mathrm{T}}\mathbf{W}\mathbf{p}$ | $I=\mathbf{z}^{\mathrm{T}}\mathbf{W}\mathbf{z}$ |
| **Local index** | | $E_i=\mathrm{diag}\,(\mathbf{Q}\mathbf{Q}^{\mathrm{T}}\mathbf{V})$ | $G_i=(\mathbf{W}\mathbf{p})_i$ | $I_i=\mathrm{diag}\,(\mathbf{z}\mathbf{z}^{\mathrm{T}}\mathbf{W})$ |
| **Scatterplot** | Scatterpoints | $\mathbf{f}=\mathbf{Q}^{\mathrm{T}}\mathbf{Q}\mathbf{W}\mathbf{Q}$ v.s. $\mathbf{Q}$ | $\mathbf{f}=\mathbf{p}^{\mathrm{T}}\mathbf{p}\mathbf{W}\mathbf{p}$ v.s. $\mathbf{p}$ | $\mathbf{f}=\mathbf{z}^{\mathrm{T}}\mathbf{z}\mathbf{W}\mathbf{z}$ v.s. $\mathbf{z}$ |

| | Trendline | $\mathbf{f}^*=\mathbf{QQ}^T\mathbf{WQ}$ v.s. $\mathbf{Q}$ | $\mathbf{f}^*=\mathbf{pp}^T\mathbf{Wp}$ v.s. $\mathbf{p}$ | $\mathbf{f}^*=\mathbf{zz}^T\mathbf{Wz}$ v.s. $\mathbf{z}$ |
|---|---|---|---|---|
| **Geographical meaning** | | Absolute correlation strength | Rescaled absolute correlation strength | Relative correlation strength |

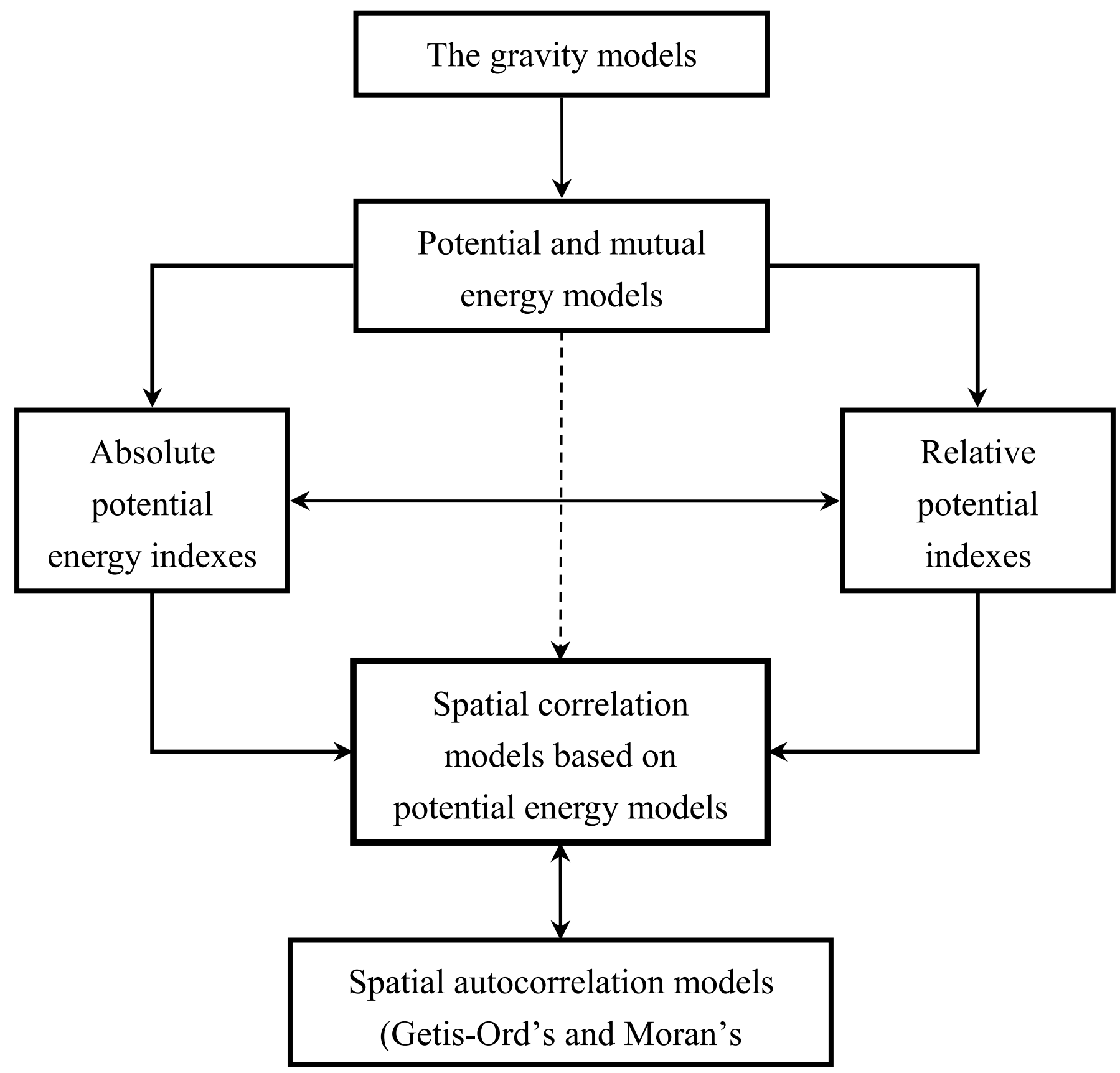

**Figure 1 An integrated analytical framework of spatial correlation based on urban potential energy models**

**Note**: A flow chart for the analytical process based on gravity models, potential energy model, potential-based spatial correlation models, and common spatial autocorrelation models

# 3 Empirical analysis

## 3.1 Study area, time, and measurements

Now, the gravity model, potential energy models and related measures, and spatial autocorrelation statistics have been integrated into the same mathematical framework. The potential energy model is actually based on fractal gravity model. It is necessary to make a case study on cities to show how to use this analytical process. In empirical work, the spatial autocorrelation analysis and the gravity modeling can complement each other. Each methodology has its own advantages and sphere of application. The spatial autocorrelation models can be employed to analyze the intra-sample correlation strength without taking into the absolute size of cities into count. The gravity-based spatial correlation model can also be used to study intra-sample correlation strength, but the effect is different. By contrast with spatial autocorrelation, the new spatial correlation analysis based on gravity model takes into consideration the sizes of cities.

The systems of cities of China can be taken as an example to make an empirical analysis of spatial correlation based on urban potential energy model. The study area covers the whole Chinese Mainland. The research subjects include the capital cities of the 31 provinces, autonomous regions, and municipalities directly under the Central Government of China (CCC). Time involves two years, that is, 2000 and 2010. The size variable ($x_i$) is city population, and the distance measure ($r_{ij}$) is interurban railway mileages. Urban population datasets come from the fifth census (in 2000) and the sixth census (in 2010), and the railway mileage data can be found in China transportation atlas. The interurban railway distances are used to construct spatial contiguity matrix, and urban population is utilized to generate size vector. Only 29 cities are really taken into account in this spatial analysis because that the cities of Haikou and Lhasa were not connected to the network of Chinese cities by railway for a long time. That is, the real size of the spatial sample is $n$=29.

## 3.2 Calculations and analysis

Using the models and measurements derived above, we can make an empirical analysis step by step. The analysis process can be divided into five stages.

**Step 1: Preparatory work**. Preparatory work includes measure selection, data acquisition, variable preprocessing, and parameter calibration. The measurement of urban size includes urban population, urbanized area, and urban Gross Domestic Product (GDP). The basic measurement of urban size are population and wealth (Dendrinos, 1992), and population reflects the first dynamics of urban evolution (Arbesman, 2012). Therefore, as indicated above, urban population is chosen as a size measure of this case. The distance measures include Euclidean distance, railway distance, highway distance, transportation cost, and travel time length. Due to the large territory of China, railway length is more suitable for measuring the distance between large cities. The significant shortcoming of energy indexes lies in dependence on dimension (quantitative units) of size variables and decay of distances. It is necessary to consider variable transform and calibration of distance decay exponent. There are various methods of value transforms for variables (Appendix). In this work, standard deviation is used to lessen the influence of quantitative units. The formula is

$$Q_i = \frac{x_i}{\sigma}, \tag{47}$$

where $x_i$ denotes the original variable, and $\sigma$ refers to population standard deviation. As a demonstrating case of newly improved analytical methodology, the distance decay exponent is set as $b$=2 for the time being. Using the formulae derived above, we can obtain the required calculation results (Table 2, Table 3).

**Table 2 The potential energy indexes and relative potential indexes of the main cities in Chinese Mainland (2000-2010)**

| City | 2000 | | | 2010 | | |
|---|---|---|---|---|---|---|
| | Size (**Q**) | Potential energy (**G**) | Relative potential (**E**) | Size (**Q**) | Potential energy (**G**) | Relative potential (**E**) |
| Beijing | 3.5696 | 0.0660 | 0.2355 | 4.0158 | 0.0671 | 0.2694 |
| Tianjin | 1.9973 | 0.0828 | 0.1655 | 2.2868 | 0.0859 | 0.1965 |
| Shijiazhuang | 0.7257 | 0.0757 | 0.0549 | 0.7119 | 0.0764 | 0.0544 |
| Taiyuan | 0.9541 | 0.0569 | 0.0542 | 0.7899 | 0.0573 | 0.0453 |
| Hohhot | 0.3725 | 0.0402 | 0.0150 | 0.3710 | 0.0405 | 0.0150 |
| Shenyang | 1.6331 | 0.0420 | 0.0686 | 1.4630 | 0.0417 | 0.0610 |
| Changchun | 1.0135 | 0.0436 | 0.0442 | 0.8495 | 0.0426 | 0.0362 |
| Harbin | 1.2987 | 0.0346 | 0.0449 | 1.2035 | 0.0334 | 0.0402 |
| Shanghai | 4.7814 | 0.0448 | 0.2142 | 4.5551 | 0.0468 | 0.2133 |
| Nanjing | 1.2957 | 0.0657 | 0.0851 | 1.4556 | 0.0654 | 0.0952 |
| Hangzhou | 0.9214 | 0.0738 | 0.0680 | 1.1391 | 0.0722 | 0.0822 |
| Hefei | 0.5498 | 0.0587 | 0.0323 | 0.7943 | 0.0584 | 0.0464 |
| Fuzhou | 0.7446 | 0.0344 | 0.0256 | 0.6908 | 0.0335 | 0.0232 |
| Nanchang | 0.6276 | 0.0562 | 0.0353 | 0.4951 | 0.0546 | 0.0270 |
| Jinan | 0.9720 | 0.0628 | 0.0611 | 0.8683 | 0.0642 | 0.0557 |
| Zhengzhou | 0.9362 | 0.0619 | 0.0580 | 0.9367 | 0.0608 | 0.0570 |
| Wuhan | 2.4655 | 0.0472 | 0.1163 | 1.8797 | 0.0468 | 0.0880 |
| Changsha | 0.7979 | 0.0522 | 0.0416 | 0.7429 | 0.0493 | 0.0366 |
| Guangzhou | 2.5818 | 0.0292 | 0.0753 | 2.3867 | 0.0284 | 0.0678 |
| Nanning | 0.5091 | 0.0302 | 0.0154 | 0.6260 | 0.0292 | 0.0183 |
| Chongqing | 2.1336 | 0.0337 | 0.0719 | 2.2390 | 0.0329 | 0.0737 |
| Chengdu | 1.4364 | 0.0349 | 0.0501 | 1.5217 | 0.0344 | 0.0523 |
| Guiyang | 0.6877 | 0.0375 | 0.0258 | 0.6276 | 0.0370 | 0.0232 |
| Kunming | 0.9316 | 0.0262 | 0.0244 | 0.8117 | 0.0258 | 0.0210 |
| Xi'an | 1.3702 | 0.0446 | 0.0611 | 1.2611 | 0.0441 | 0.0556 |
| Lanzhou | 0.6823 | 0.0346 | 0.0236 | 0.6148 | 0.0342 | 0.0210 |
| Xining | 0.3212 | 0.0331 | 0.0106 | 0.2946 | 0.0324 | 0.0095 |
| Yinchuan | 0.2174 | 0.0349 | 0.0076 | 0.2893 | 0.0343 | 0.0099 |
| Urumqi | 0.6494 | 0.0156 | 0.0101 | 0.7199 | 0.0154 | 0.0111 |
| **Mean** | **1.2820** | **0.0467** | **0.0619** | **1.2635** | **0.0464** | **0.0623** |
| **Stdev** | **1.0000** | **0.0163** | **0.0554** | **1.0000** | **0.0167** | **0.0612** |

**Note**: The "standardized" urban population size $Q$ is generated by using equation (47).

**Step 2: Calculation of relative potential indexes**. Relative potential indexes can reflect the relative geographical advantages of a city, mainly the locational advantages, especially the traffic reachability. The sum of local relative potential indices is equal to the global relative potential index. Global relative potential index is chiefly used for comparison between different urban systems. For the spatial advantages such as accessibility within an urban system, it is sufficient to compare the local relative potential indexes of different cities. For the provincial capital cities in Chinese Mainland, the relative potential indexes are ranked as follows (Figure 2). From 2000 to 2010, there have been some changes in the location advantages of different cities, but on the whole, the changes are not significant. Shanghai's position has advanced, while Wuhan's position is relatively backward.

**Table 3 The inner product and outer product of weighted variables of the main cities in Chinese Mainland (2000-2010)**

| **City** | **2000** | | | **2010** | | |
|---|---|---|---|---|---|---|
| | **Q** | **$Q^TQWQ$** | **$QQ^TWQ$** | **Q** | **$Q^TQWQ$** | **$QQ^TWQ$** |
| Beijing | 3.5696 | 5.0572 | 6.4111 | 4.0158 | 5.0520 | 7.2527 |
| Tianjin | 1.9973 | 6.3505 | 3.5872 | 2.2868 | 6.4709 | 4.1300 |
| Shijiazhuang | 0.7257 | 5.8014 | 1.3033 | 0.7119 | 5.7493 | 1.2856 |
| Taiyuan | 0.9541 | 4.3585 | 1.7136 | 0.7899 | 4.3172 | 1.4266 |
| Hohhot | 0.3725 | 3.0845 | 0.6690 | 0.3710 | 3.0511 | 0.6699 |
| Shenyang | 1.6331 | 3.2200 | 2.9332 | 1.4630 | 3.1411 | 2.6423 |
| Changchun | 1.0135 | 3.3406 | 1.8203 | 0.8495 | 3.2048 | 1.5342 |
| Harbin | 1.2987 | 2.6520 | 2.3325 | 1.2035 | 2.5132 | 2.1735 |
| Shanghai | 4.7814 | 3.4337 | 8.5876 | 4.5551 | 3.5254 | 8.2266 |
| Nanjing | 1.2957 | 5.0330 | 2.3272 | 1.4556 | 4.9243 | 2.6288 |
| Hangzhou | 0.9214 | 5.6582 | 1.6549 | 1.1391 | 5.4331 | 2.0572 |
| Hefei | 0.5498 | 4.5016 | 0.9875 | 0.7943 | 4.3986 | 1.4346 |
| Fuzhou | 0.7446 | 2.6351 | 1.3374 | 0.6908 | 2.5257 | 1.2475 |
| Nanchang | 0.6276 | 4.3083 | 1.1272 | 0.4951 | 4.1109 | 0.8942 |
| Jinan | 0.9720 | 4.8155 | 1.7458 | 0.8683 | 4.8321 | 1.5682 |
| Zhengzhou | 0.9362 | 4.7462 | 1.6815 | 0.9367 | 4.5788 | 1.6918 |

| Wuhan | 2.4655 | 3.6171 | 4.4282 | 1.8797 | 3.5244 | 3.3948 |
|---|---|---|---|---|---|---|
| Changsha | 0.7979 | 3.9986 | 1.4331 | 0.7429 | 3.7146 | 1.3416 |
| Guangzhou | 2.5818 | 2.2357 | 4.6370 | 2.3867 | 2.1401 | 4.3104 |
| Nanning | 0.5091 | 2.3148 | 0.9145 | 0.6260 | 2.2023 | 1.1306 |
| Chongqing | 2.1336 | 2.5837 | 3.8321 | 2.2390 | 2.4773 | 4.0436 |
| Chengdu | 1.4364 | 2.6730 | 2.5799 | 1.5217 | 2.5871 | 2.7482 |
| Guiyang | 0.6877 | 2.8718 | 1.2351 | 0.6276 | 2.7844 | 1.1334 |
| Kunming | 0.9316 | 2.0083 | 1.6733 | 0.8117 | 1.9435 | 1.4659 |
| Xi'an | 1.3702 | 3.4196 | 2.4609 | 1.2611 | 3.3212 | 2.2777 |
| Lanzhou | 0.6823 | 2.6543 | 1.2255 | 0.6148 | 2.5762 | 1.1104 |
| Xining | 0.3212 | 2.5369 | 0.5768 | 0.2946 | 2.4371 | 0.5321 |
| Yinchuan | 0.2174 | 2.6728 | 0.3904 | 0.2893 | 2.5858 | 0.5224 |
| Urumqi | 0.6494 | 1.1969 | 1.1664 | 0.7199 | 1.1575 | 1.3001 |
| **Mean** | **1.2820** | **3.5786** | **2.3025** | **1.2635** | **3.4924** | **2.2819** |
| **Stdev** | **1.0000** | **1.2477** | **1.7961** | **1.0000** | **1.2583** | **1.8060** |

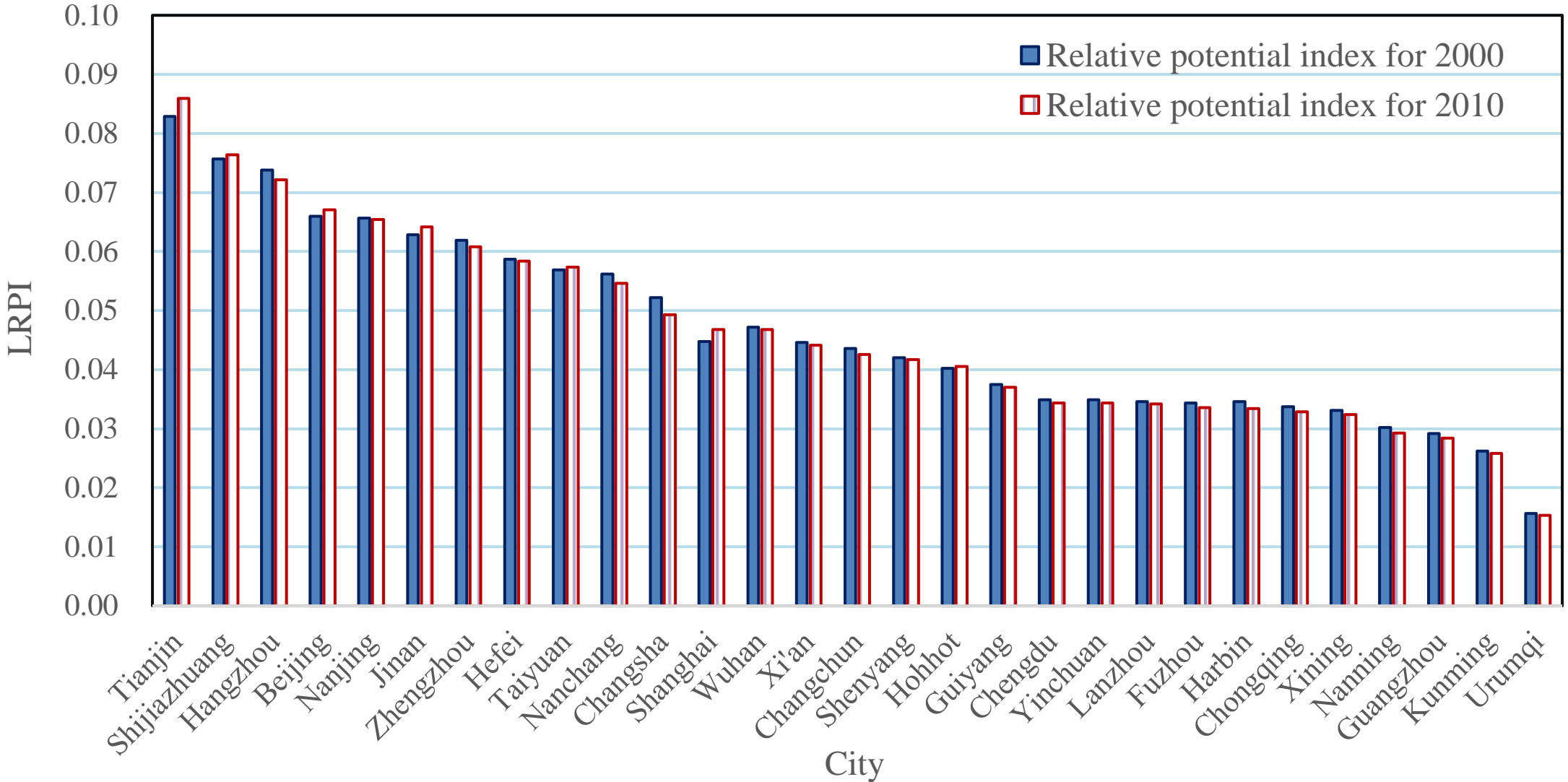

**Figure 2 The local relative potential indexes of the main cities in Mainland China (2000 & 2010)**

**Note:** The histogram chart is generated using the data in Table 2. Sort based on numerical calculations of 2010.

**Step 3: Calculation of absolute potential energy indexes**. The relative potential index is mainly utilized to examine a city from the perspective of its location in a network of cities, without considering its size advantages of a city itself. In the absolute potential energy index, the impact of a city's own size is further considered on the basis of the relative potential index. Therefore, absolute potential energy index can reflect a city's absolute advantage in the urban system. For provincial capital cities in Chinese Mainland, the ranking of absolute potential energy index values are shown as below (Figure 3). From 2000 to 2010, the absolute potential energy indexes have some changes, but in a mass, the changes are also not significant. Nanjing's position has advanced, while Wuhan's position is relatively backward; Hefei's position has advanced, while Taiyuan's position is relatively backward.

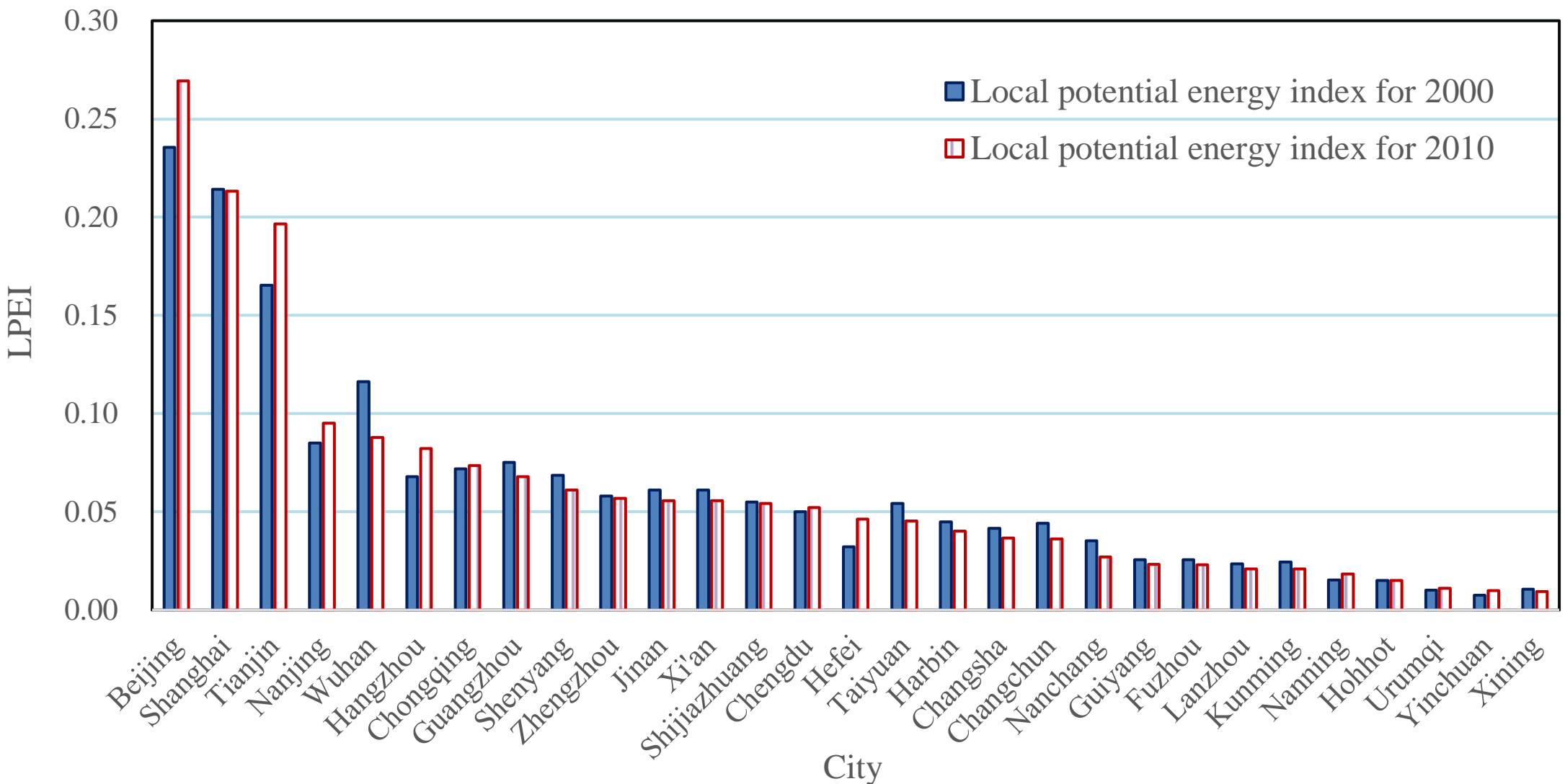

**Figure 3 The local potential energy indexes of the main cities in Mainland China (2000 & 2010)**

**Note:** The histogram chart is generated using the data in Table 2. Sort based on numerical calculations of 2010.

**Step 4: Spatial correlation scatterplots**. Three sets of spatial correlation scatterplots can be employed to make visually spatial analysis. The first scatter plot is the absolute potential energy scatter plot, as the slope of the trend line gives the global absolute potential energy index value. This potential scatterplot comes from analogy with Moran's scatterplot in spatial autocorrelation analysis. For the absolute potential energy scatterplot, the abscissa axis ($x$-axis) can be characterized by standardized or normalized size variable, $\mathbf{Q}$, and vertical axis ($y$-axis) can be described with corresponding inner product weight variable, $\lambda\mathbf{WQ}$, where $\lambda=\mathbf{Q}^{\mathrm{T}}\mathbf{Q}$ (Figure 4). If the intercept is fixed to 0, then the slope of the trend line gives the global absolute potential energy index, i.e. the value of $E$. If the outer product weighted variables, $\mathbf{QQ}^{\mathrm{T}}\mathbf{WQ}$, is added to the vertical axis, it will give another trend line, which is overlapped with the first trend line. Cities below the trend line have lower absolute potential energy relative to their size, including Beijing, Shanghai, Chongqing, and Guangzhou. Cities above the trend line have higher absolute potential energy relative to their size, including Tianjin, Hangzhou, Shijiazhuang, and Nanjing. The absolute potential energy of cities on or near the trend line roughly matches their size, including Chengdu, Urumqi, and Harbin. The more to the right of the scatter plot, the larger the city size; the higher up the scatter plot, the higher the value of the urban relative potential index. With the help of the average values of city size variable (**Q**) and the average value of relative potential index (**WQ**) (see Table 3 for the means), the major cities in Chinese Mainland can be divided into four types (see Table 4 for the results). It can be seen that, from 2000 to 2010, the status of Shanghai increased, while the status of Harbin decreased. The status of other cities has not significantly changed in the classification pattern.

**Table 4 Classification of cities in Chinese Mainland based on absolute potential scatterplot (2000-2010)**

| Year | Potential | Smaller size | Bigger size |
|---|---|---|---|
| **2000** | Bigger potential | Shijiazhuang, Hangzhou, Jinan, Zhengzhou, Hefei, Taiyuan, Nanchang, Changsha (8 cities) | Tianjin, Beijing, Nanjing, Wuhan (4 cities) |
| | Smaller potential | Changchun, Kunming, Fuzhou, Guiyang, Lanzhou, Urumqi, Nanning, Hohhot, Xining, Yinchuan (10 cities) | Shanghai, Guangzhou, Chongqing, Shenyang, Chengdu, Xi'an, Harbin (7 cities) |
| **2010** | Bigger potential | Shijiazhuang, Hangzhou, Jinan, Zhengzhou, Hefei, Taiyuan, Nanchang, Changsha (8 cities) | Tianjin, Beijing, Nanjing, Shanghai, Wuhan (5 cities) |
| | Smaller potential | Harbin, Changchun, Kunming, Urumqi, Fuzhou, Guiyang, Nanning, Lanzhou, Hohhot, Xining, Yinchuan (11 cities) | Guangzhou, Chongqing, Shenyang, Chengdu, Xi'an (5 cities) |

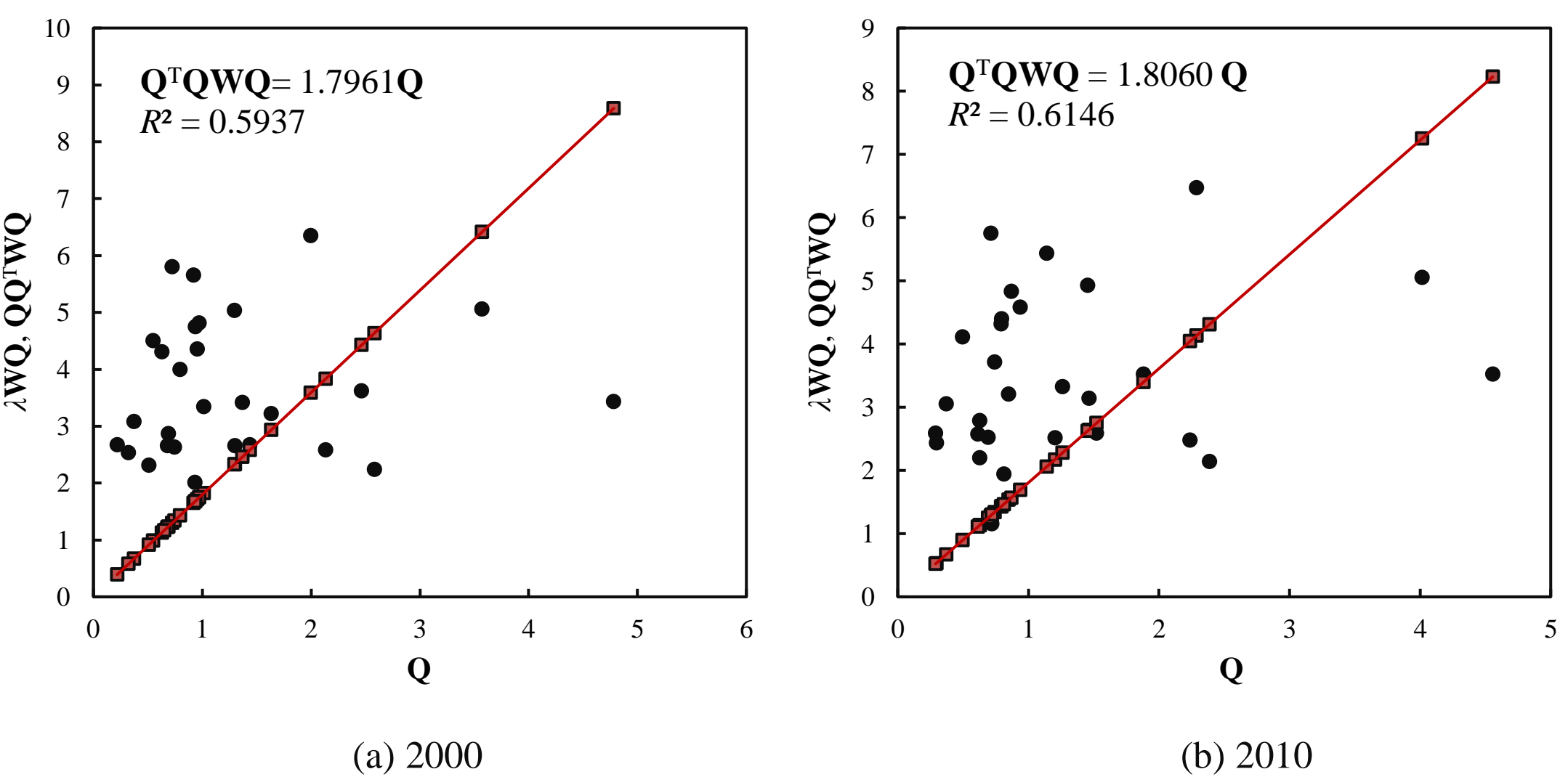

(a) 2000 (b) 2010

**Figure 4 Two absolute potential scatterplots based on gravity model for the main cities of China (2000 & 2010)**

(**Note**: The scatterplot is generated using the data in Table 3. The black dots are given by the inner product equation of spatial correlation, $\mathbf{Q}^{\mathrm{T}}\mathbf{QWQ}=E\mathbf{Q}$, representing scattered points; the square points are given by the outer product equation of spatial correlation, $\mathbf{QQ}^{\mathrm{T}}\mathbf{WQ}=E\mathbf{Q}$, being completely consistent with the trend line.)

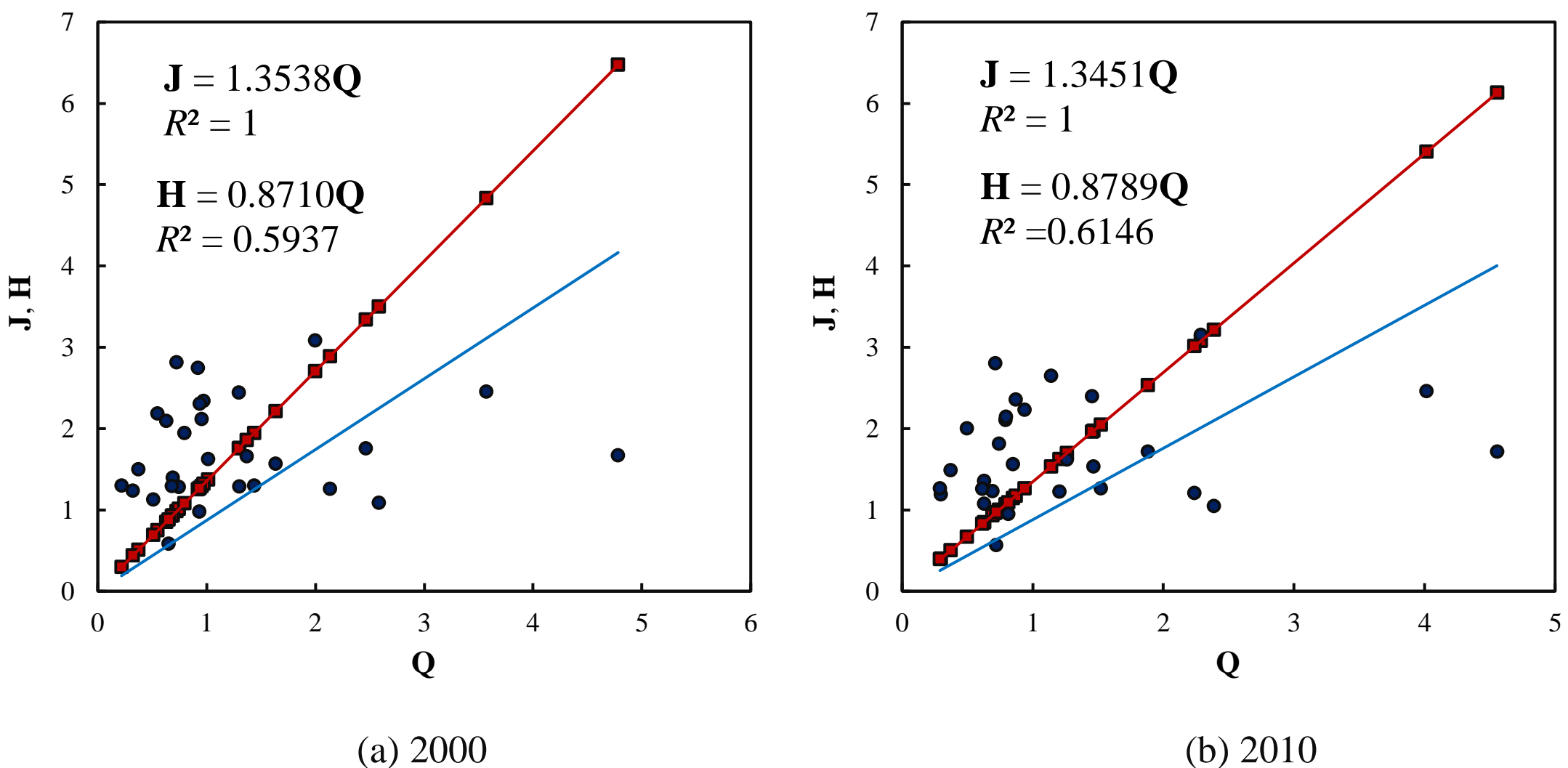

**Figure 5 Two relative potential scatterplots based on gravity model for the main cities of China (2000 & 2010) 2023-05-23**

(**Note**: The scatterplot is generated using the data in Table 2. The black dots are given by the relationship between **Q** and **H**, representing scattered points; the square points are given by the relationship between **Q** and **J**, being completely consistent with the trend line with a slope equal to the global relative potential index. The relationship between **Q** and **H** can give another trend line, which is equivalent to the trend line in Figure 4.)

The second scatter plot is the relative potential energy scatter plot because the slope of the trend line gives the value of the global relative potential energy index value. The kind of scatterplot is also an innovation of this work. For the relative potential energy scatterplot, the $x$-axis is characterized by standardized or normalized size variable, **Q**, and the $y$-axis is described with **J** and **H** values (Figure 5). The relation between **Q** and **H** displays a group of randomly distributed data points. In contrast, the relation between **Q** and **J** takes on a trend line, the slope of which equals the global relative potential index. The relative potential energy (traffic accessibility) of the cities below the trend line is lower relative to their size. This type of cities includes Beijing, Shanghai, Chongqing, and Guangzhou. The relative potential energy of the cities above the trend line is higher relative to their size. This type of cities includes Shijiazhuang, Hangzhou, Nanjing, and Zhengzhou. The relative potential energy of the cities on or near the trend line roughly matches their size. In 2000, this type of city was not very obvious. By 2010, three cities bear traffic reachability matching their own sizes, including Tianjin, Xi’an, and Kunming. A trend line with an intercept of 0 can also be given based on the relationship between **Q** and **H**, but its slope is neither equal to the absolute potential energy index nor equal to the relative potential energy index. However, the result of matching this trend line with a scattered points is equivalent to an absolute potential energy scatter plot. This means that the relative potential scatter plot actually contains information about the absolute potential scatter plot. So, urban classification based on the absolute potential energy scatter plot can also be obtained from the relative potential energy scatter plot (Table 4).

The third scatter plot is the absolute-relative potential energy scatter plot because it is constructed by means of absolute and relative potential energy indexes. The kind of scatterplot is also an innovation of this work. For the absolute-relative potential energy scatterplot, the abscissa axis is represented by local relative potential index, **G**, and vertical axis is reflected by corresponding local potential energy index, **E** (Figure 6). Horizontal axis variables can be replaced by inner product weight variable, $\lambda$**WQ**. The more to the right of the scatter plot, the higher the relative potential index value; the higher up the scatter plot, the higher the value of the absolute potential energy index. With the help of the average values of local relative potential index (**G**) and the average value of local absolute potential index (**E**) (see Table 2 for the means), the major cities of Chinese Mainland can also be divided into four categories (see Table 5 for the results). Overall, from 2000 to 2010, Shanghai's position improved, while Shenyang's position declined. The status of other cities has not significantly changed in the classification pattern.

**Table 5 Classification of cities in Chinese Mainland based on potential-mutual energy scatterplot (2000-2010)**

| Year | Potential | Smaller potential energy | Bigger potential energy |
|---|---|---|---|
| **2000** | Bigger mutual energy | Shanghai, Guangzhou, Chongqing, Shenyang (4 cities) | Beijing, Tianjin, Wuhan, Nanjing, Hangzhou (5 cities) |
| | Smaller mutual energy | Xi'an, Changchun, Hohhot, Guiyang, Chengdu, Yinchuan, Lanzhou, Harbin, Fuzhou, Xining, Nanning, Kunming, Urumqi (12 cities) | Shijiazhuang, Jinan, Zhengzhou, Hefei, Taiyuan, Nanchang, Changsha (7 cities) |
| **2010** | Bigger mutual energy | Guangzhou, Chongqing (2 cities) | Beijing, Shanghai, Tianjin, Wuhan, Nanjing, Hangzhou (6) |
| | Smaller mutual energy | Xi'an, Changchun, Shenyang, Hohhot, Guiyang, Chengdu, Yinchuan, Lanzhou, Fuzhou, Harbin, Xining, Nanning, Kunming, Urumqi (14 cities) | Shijiazhuang, Jinan, Zhengzhou, Hefei, Taiyuan, Nanchang, Changsha (7 cities) |

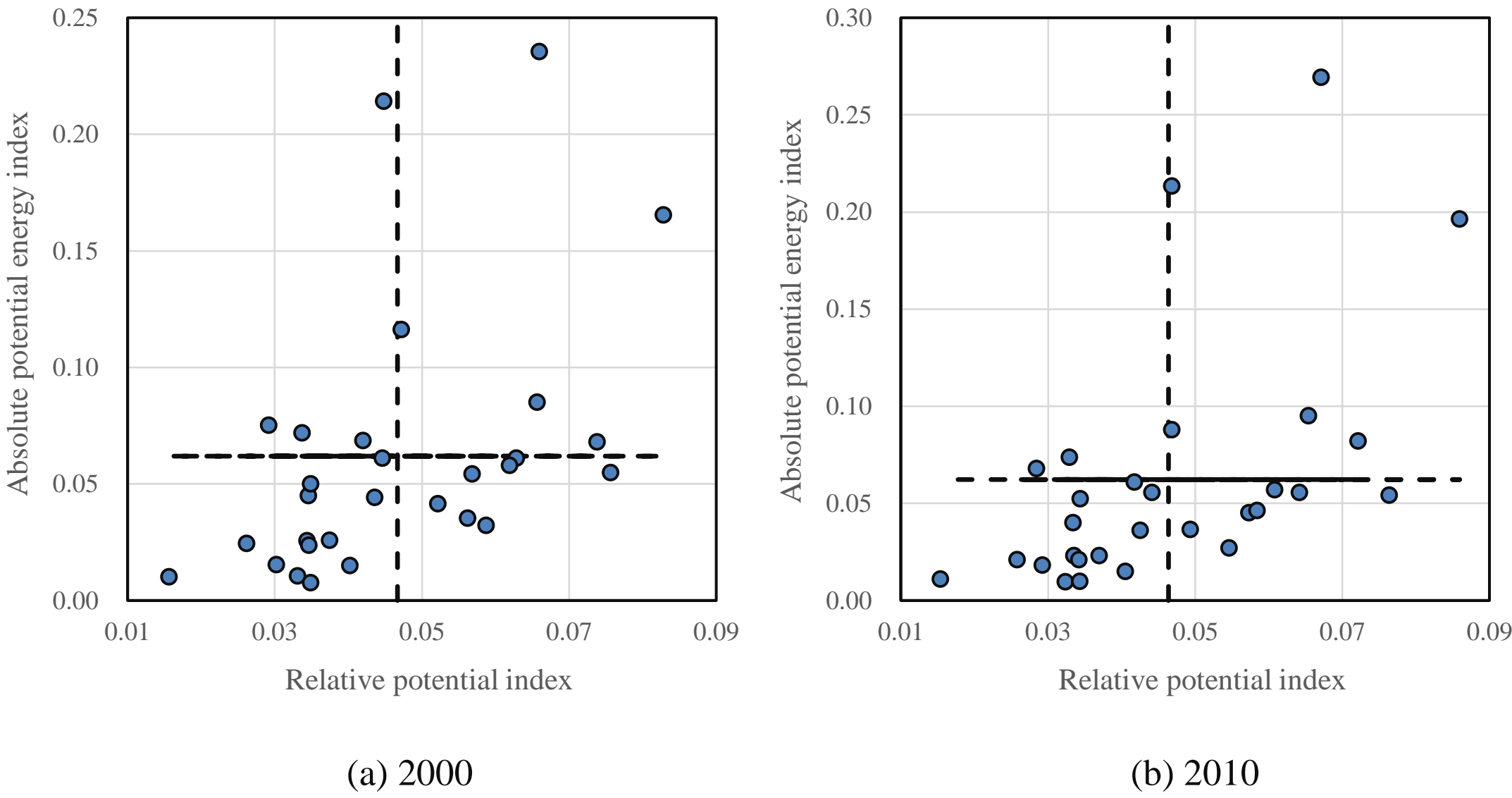

(a) 2000 (b) 2010

**Figure 6 Two absolute-relative potential energy scatterplots based on gravity model for the main cities of China (2000 & 2010)**

(**Note**: The scatterplot is generated using the data in Table 2. Dots represent local relative potential indexes and absolute potential energy indexes. The vertical line represents the average line of local relative potential indexes, and the horizontal line represents the average line of local absolute potential energy indexes.)

**Step 5: Parameter calibration by idea from fractal dimension**. The usual spatial autocorrelation analysis does not involve the distance decay index. The distance decay exponent is often defaulted to be $b$=2, and accordingly, the spatial weight function takes the reciprocal of the distance. In this case, spatial autocorrelation does not involve parameter calibration issues. Now we know that the spatial autocorrelation function can be derived from the gravity model, and the distance decay exponent is actually a fractal parameter. Therefore, the spatial weight function is in fact an inverse power law such as

$$v_{ij} = \frac{1}{r_{ij}^{b}} = \frac{1}{r_{ij}^{qD-1}} , \tag{48}$$

where $b=qD$, $q$ refers to Zipf's exponent of rank-size distribution of cities, and $D$ denotes the fractal dimension of central place systems or networks of cities (Chen, 2015). Generally speaking, $q$=1, and $D$ varies from 1 to 2. However, in empirical calculation, the fractal dimension value may exceed a reasonable range. That is, the fractal parameter value may be less than 1 or greater than 3. Assuming that the definition of variables remains unchanged, that is, it is based on equation (47), we can investigate the dependence of absolute and relative potential energy index values on fractal parameter values. The calculation results show that, as the fractal parameter value increases, the spatial correlation index values also increases (Figure 7). Compared to the relative potential indexes, the absolute potential energy indexes has a more significant dependence on fractal parameters.

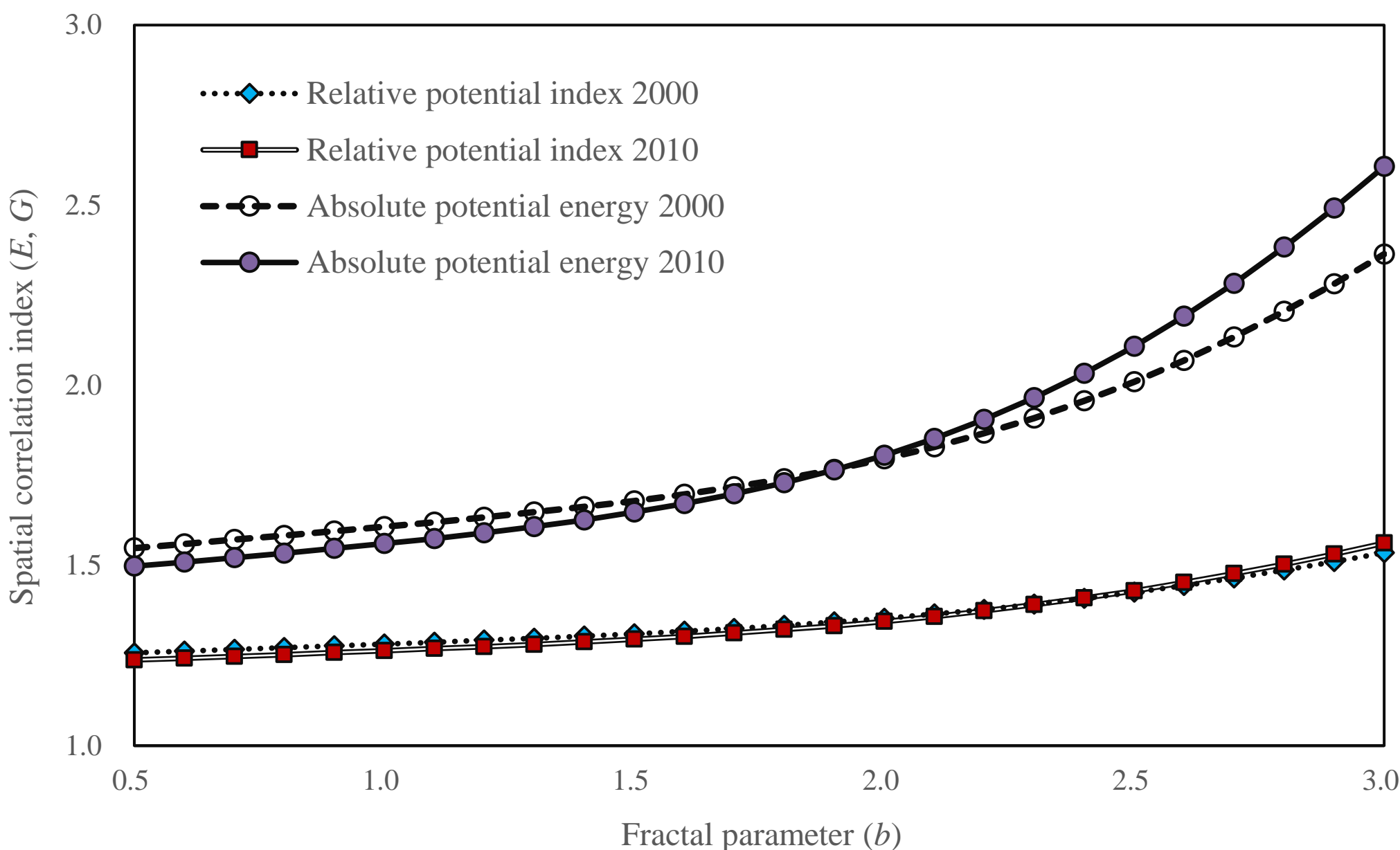

**Figure 7 Dependence of absolute and relative potential energy indexes of the main Chinese cities on fractal dimension parameter values (2000 & 2010)**

# 4 Discussion

A new methodology of spatial correlation analysis has been developed by means of concepts and formulae of gravity and potential energy in social physics and human geography. The potential energy models and measures are derived from the gravity models, which can be derived from the spatial interaction models (Chen, 2015). The geographical gravity model is actually a fractal model (Chen, 2023). In this sense, the spatial autocorrelation analysis can be associated with spatial interaction analysis. On the other hand, spatial autocorrelation measures can be derived from new potential energy models. The potential-based correlation models bear an analogy with the spatial autocorrelation models based on Getis-Ord's indexes and Moran's indexes. In this sense, spatial autocorrelation analysis and spatial interaction analysis reach the same goal by different routes in geographical spatial modeling. In particular, fractal concept can be employed to improve the method of geo-spatial analysis.

The analytical framework of the new methodology comprises two models: one is the absolute potential energy model, and the other, relative potential models. Each model comprises two basic measurements and a scatterplot (Table 6). The two measurements are the global index (global absolute potential energy index, global relative potential index) and local indexes (local absolute potential energy index, local relative potential index). The global index reflects the sum of the correlation strength between any two cities, while the local index reflects the sum of the correlation strength between a city and all other cities. Thus the global index is a correlation measurement of the whole system of cities, and local index is a correlation measurement of a city within the system of cities. Where the potential-based model of cities is concerned, the contribution to the global correlation strength comes from two respects: one is city size distribution, and the other is network structure; the contribution to the local correlation strength also results from two aspects: one is the size of city, and the other, the geographical location of the city in the network of cities. The larger and the closer to the center of gravity in a geographical region a city is, the larger the local absolute potential energy index will be.

**Table 6 The components of the spatial correlation analysis based on potential energy and the geographical meaning of the potential energy indexes**

| Item | | Absolute potential energy model | Relative potential model |
|---|---|---|---|
| Global index | Formula | $E=\mathbf{Q}^{\mathrm{T}}\mathbf{W}\mathbf{Q}$ | $S=\sum(\mathbf{WQ})_i$ |
| | Meaning | Two-order size distribution and network structure | One-order size distribution and network structure |
| Local index | Formula | $E_i=\mathrm{diag}(\mathbf{Q}\mathbf{Q}^{\mathrm{T}}\mathbf{W})$ | $U_i=(\mathbf{WQ})_i$ |
| | Meaning | Two-order size and geographical location | One-order size and geographical location |
| Scatterplot | Scatterpoints | $\mathbf{f}=\mathbf{Q}^{\mathrm{T}}\mathbf{QWQ}$ v.s. $\mathbf{Q}$ | $H_i$ v.s. $Q_i$ |
| | Trend line | $\mathbf{f}^{*}=\mathbf{QQ}^{\mathrm{T}}\mathbf{WQ}$ v.s. $\mathbf{Q}$ | $J_i$ v.s. $Q_i$ |
| Geographical meaning | | Absolute importance based on size and location | Relative importance based on location in a network |

**Note**: The symbol "diag" denotes "take diagonal elements from a square matrix".

The spatial correlation models proposed in this paper is based on the concepts from potentials in urban geography and social physics. The potential models resulted from the traditional gravity models, but the mathematical expressions of gravity models varied in social sciences. In fact, there two types of basic gravity models. One is based on inverse power law distance decay, as shown by equation (1), and the other is based on negative exponential distance decay. The gravity model based on negative exponential decay can be expressed as

$$F_{ij} = KQ_iQ_je^{-r_{ij}/r_0} \,, \tag{49}$$

where $r_0$ is a scale parameter of spatial interaction, and the rest notation is the same as in equation (1). Accordingly, the absolute potential energy of city $i$ and city $j$ can be given by

$$E_{ij} = F_{ij}r_{ij} = KQ_iQ_jr_{ij}e^{-r_{ij}/r_0} = Kv_{ij}e^{-r_{ij}/r_0} \,, \tag{50}$$

where

$$v_{ij} = r_{ij}e^{-r_{ij}/r_0} \,, \tag{51}$$

is a gamma function, representing a spatial weight function. Regarding the gravity coefficient as a scaling factor and defining

$$K = \frac{1}{V_0} = 1/\sum_{i=1}^{n}\sum_{j=1}^{n}v_{ij} = 1/\sum_{i=1}^{n}\sum_{j=1}^{n}r_{ij}e^{-r_{ij}/r_0} \,, \tag{52}$$

we can derive a set of absolute and relative potential measurements based on the negative exponential decay function. It can be proved that the scale parameter equals half average distance in a 2-dimensional space. In practice, the scale parameter can be estimated with the mean of the distances between the $n$ cities, $\bar{r}$, and we have

$$\hat{r}_0 = \frac{\bar{r}}{2} = \frac{1}{n(n+1)}\sum_{i=1}^{n}\sum_{j=1}^{i}r_{ij} = \frac{1}{2n(n+1)}\sum_{i=1}^{n}\sum_{j=1}^{n}r_{ij} \,, \tag{53}$$

where $\bar{r}$ is the average distance, and $\hat{r}_0$ is the estimated value of $r_0$. If the absolute potential energy model is based on the equation (49), the spatial contiguity function can be defined as below:

$$v_{ij} = \begin{cases} r_{ij}^{D-1}\exp(-r_{ij}/r_0), & i \neq j \\ 0, & i = j \end{cases} \,, \tag{54}$$

where $D$ is a quasi-fractal dimension, coming between 1 and 3. Using equation (54), we can generate a spatial contiguity matrix, which leads to a weight matrix.

The original form of the distance-decay function of the gravity model is a power function. However, the inverse power-law decay function was once replaced by the negative exponential decay function. The reason for this is that the dimension meaning of the distance exponent $b$ could not be interpreted by using the ideas from Euclidean geometry, and especially, the model seemed to be not derivable from general principle. The advantages of the exponential-based gravity model are as follows: first, it is independent of dimension; second, its underlying rationale is clear because it is derivable from the principle of entropy maximization (Haggett *et al*, 1977; Haynes, 1975; Wilson, 1968; Wilson, 1970). However, the exponential-based gravity gave rise to new problems because that the exponential decay function suggests locality or localization rather than action at a distance, which contradicts the first law of geography (Chen, 2015). On the other hand, the original gravity model based on power law decay can be derived from the principles of allometric scaling and fractal, and the distance exponent can be interpreted with the concept of fractal dimension (Chen, 2023). Despite this, the exponential-based gravity model can be employed to make an analysis of spatial interaction for smaller regions or simpler systems.

The novelty of this work mainly lies in three aspects. First, the spatial interaction modeling is associated with spatial autocorrelation. Spatial autocorrelation coefficients, including Moran's index and Getis-Ord's index, are derived from the gravity models. Second, the concept of fractal dimension in the weight function is revealed. The gravity model based on power law is a fractal model. Third, the methods of spatial analysis is developed. Based on absolute and relative potential energy concepts, a new analytical framework of geographical spatial analysis is proposed. Moreover, matrix multiplication is introduced into the correlation model to simplify the mathematical expressions, and a set of new measurements and three types of scatterplots are presented. The shortcomings of this study are chiefly as follows. First, scaling process is not taken into account. Geographical spatial measurement often depend on scales, which leads to a spatial scaling process, but this process has not been considered for the time being. Second, the new method is based on univariate rather than multivariate analysis. Although multiple elements of a geographical system have been considered, multiple variables have not yet been introduced into the analytical framework.

# 5 Conclusions

The new analytical processes of spatial correlation analysis based on the gravity model developed in this paper is useful in both theory, method, and practice. As far as theory is concerned, it unifies three important elements involving spatial complexity, that is, spatial dimension, time lag (spatial displacement), and interaction. Where method and practice are concerned, it can integrate spatial autocorrelation analysis and spatial interaction modeling into the same logic framework. Based on the theoretical derivation and empirical study, the main conclusions can be drawn as follows. *First,* *the concepts of human force, energy, and potential can be organized to form a new framework of spatial correlation analysis*. The basic measurements include the global absolute potential energy (GAPE) index, local absolute potential energy (LAPE) index, global relative potential (GRP) index, and local relative potential (LRP) index. GAPE index indicates the total strength of spatial interaction of all cities, LAPE index indicates the strength of interaction between a city and all other cities, GRP index indicates total accessibility of an urban system, and LRP index indicates the accessibility of a city in the urban system. The size-potential scatterplot can be used to reflect the local spatial autocorrelation and spatial accessibility visually, and the absolute-relative potential energy scatterplot can be employed to reflect the absolute and relative advantages of geographical elements such as cities. Using the scatterplots, we can classify geographical elements in a region according to spatial correlation and interaction. *Second, the main measures of spatial autocorrelation analysis can be derived from the gravity models.* The spatial autocorrelation and spatial interaction represent the most important methods of spatial analysis in geography. The fractal gravity model can be derived from the spatial interaction model by means of allometric scaling relations. The spatial correlation model based on the potential energy was derived from the fractal gravity model, and thus can be associated with the spatial autocorrelation model. Moran's index was proved to be a standardized potential energy index, while Getis-Ord's indexes was proved to be normalized potential energy indexes. Through variable transformation, we can obtain different types of spatial autocorrelation indexes, including new spatial statistic measures. So, the mathematical links between spatial autocorrelation analysis and spatial interaction modeling can be brought to light by new spatial correlation models. *Third, spatial weight matrix is associated with fractal dimension*. The geographical gravity model based on inverse power law decay is actually a fractal model, and the distance decay exponent cannot be interpreted by Euclidean geometry. It was proved to be a fractal parameter of geographical elements and networks of geographical elements. Since the spatial autocorrelation indexes can be derived from the gravity model, the distance decay exponent as a fractal parameter is naturally introduced into spatial contiguity matrix by spatial weight function. In contrast, if the gravity model is based on negative exponential decay, the corresponding spatial weight function of spatial autocorrelation was proved to be a gamma function. The gamma decay model can be generalized to a quasi-fractal model. The value of fractal parameter influence calculation results of spatial autocorrelation indexes. The relationship between spatial autocorrelation and fractal dimension needs further exploration in future.

**Acknowledgement**

This research was sponsored by the National Natural Science Foundation of China (Grant No. 42171192). The support is gratefully acknowledged.